\documentclass[12pt,preprint]{aastex}
\usepackage{natbib}
\usepackage{amssymb,amsmath} 

\shorttitle{ GRB111228A: Multi-wavelength observations and Implications }
\shortauthors{Xin et al. }

\begin{document}

\title{Multi-Wavelength Observations of GRB 111228A and Implications for the Fireball and its environment}

\author{Li-Ping Xin \altaffilmark{1,2, 3},Yuan-Zhu Wang\altaffilmark{2, 3}, Ting-Ting Lin\altaffilmark{2, 3}, En-Wei Liang \altaffilmark{2,3}, Hou-Jun L\"{u} \altaffilmark{2,3}, Shu-Qing Zhong\altaffilmark{2, 3}, Yuji Urata\altaffilmark{4,5}, Xiao-Hong Zhao\altaffilmark{6,7}, Chao Wu\altaffilmark{1}, Jian-Yan Wei\altaffilmark{1}, Kui-Yun Huang\altaffilmark{8}, Yu-Lei Qiu\altaffilmark{1}, Jin-Song Deng\altaffilmark{1}}
\altaffiltext{1}{Key Laboratory of Space Astronomy and Technology, National Astronomical Observatories, Chinese Academy of Sciences,China; xlp@nao.cas.cn; lew@gxu.edu.cn}
  \altaffiltext{2}{GXU-NAOC Center for Astrophysics and Space Sciences, Department of Physics, Guangxi University, Nanning 530004, China}
  \altaffiltext{3}{Guangxi Key Laboratory for the Relativistic Astrophysics, Nanning 530004, China}
  \altaffiltext{4}{Institute of Astronomy, National Central University, Chung-Li 32054, Taiwan}
  \altaffiltext{5}{Academia Sinica Institute of Astronomy and Astrophysics, Taipei 106, Taiwan}
  \altaffiltext{6}{Yunnan Observatories, Chinese Academy of Sciences, P.O. Box 110, 650011 Kunming, China}
  \altaffiltext{7}{Key Laboratory for the Structure and Evolution of Celestial Bodies, Chinese Academy of Sciences, P.O. Box 110, 650011 Kunming, China}
  \altaffiltext{8}{Department of Mathematics and Science, National Taiwan Normal University, Lin-kou District, New Taipei City 24449, Taiwan}

\begin{abstract}
Observations of very early multi-wavelength afterglows are critical to reveal the properties of the radiating fireball and its environment as well as the central engine of gamma-ray bursts (GRBs). We report our optical observations of GRB 111228A from 95 sec to about 50 hours after the burst trigger and investigate its properties of the prompt gamma-rays and the ambient medium using our data and the data observed with {\em Swift} and {\em Fermi} missions. Our joint optical and X-ray spectral fits to the afterglow data show that the ambient medium features as low dust-to-gas ratio. Incorporating the energy injection effect, our best fit to the afterglow lightcurves with the standard afterglow model via the Markov Chain Monte Carlo (MCMC) technique shows that $\epsilon_e=(6.9\pm 0.3)\times 10^{-2}$, $\epsilon_B=(7.73\pm 0.62)\times 10^{-6}$, $E_{\rm K}=(6.32\pm 0.86)\times 10^{53}\rm  erg$, $n=0.100\pm 0.014$ cm$^{-3}$. The low medium density likely implies that the afterglow jet may be in a halo or in a hot ISM. Achromatic shallow decay segment observed in the optical and X-ray bands is well explained with the long-lasting energy injection from the central engine, which would be a magnetar with a period of about 1.92 ms inferred from the data. The $E_p$ of its time-integrated prompt gamma-ray spectrum is $\sim 26$ KeV. Using the initial Lorentz factor ($\Gamma_0=476^{+225}_{-237}$) derived from our afterglow model fit, it is found that GRB 111228A satisfies the $L_{\rm iso}-E_{\rm p,z}-\Gamma_0$ relation and bridges the typical GRBs and low luminosity GRBs in this relation.

\end{abstract}


\section{Introduction}
Afterglows of gamma-ray bursts (GRBs) are generally believed to be produced by a relativistic jet interacting with the surrounding medium (e.g., M\'{e}sz\'{a}ros \& Rees 1997; Sari et al. 1998; Kumar \& Zhang 2015). Very early multi-wavelength afterglows are critical to reveal the properties of the radiating fireball and its environment as well as the central engine of GRBs. With promptly slewing and precisely locating capacity of the X-ray telescope (XRT) on board the {\em Swift} mission, very early X-ray and optical afterglows were obtained with XRT and ground-based optical telescopes. The early X-ray afterglows are usually dominated by erratic X-ray flares and the tail emission of prompt gamma-rays being due to the arrival time delay of photons in high latitude of the radiating fireball (Zhang et al. 2006, 2007; O'Brien et al. 2006; Liang et al. 2006). The flares and prompt tail emission are usually not seen in the early optical afterglow data (Li et al. 2012; Wang et al. 2013). About one-third well-sampled optical afterglow lightcurves show a clear smooth bump (Li et al. 2012). It may be attributed to deceleration of the fireball by the ambient medium (see Rees \& M{\'{e}sz\'{a}ros 1992, M{\'{e}sz\'{a}ros \& Rees 1993 and Sari \& Piran 1999 for the thin shell case;  Kobayashi et al. 1999 and Kobayashi \& Zhang 2007 for the thick shell
case). This may give the most robust estimate to initial Lorentz factor ($\Gamma_0$) of the fireball since the deceleration time weakly depends on other model parameters than $\Gamma_0$ (e.g., Molinari et al. 2007; Melandri et al. 2010; Xin et al., 2015). With a sample of GRBs with detection of an optical afterglow onset bump, Liang et al. (2010) derived their $\Gamma_0$ and found a tight $\Gamma_0-E_{\rm iso}$ relation (see also a $\Gamma_0-L_{\rm iso}$ relation in Lu et al. 2012). Furthermore, by incorporating the peak energy of the $\nu f_\nu$ spectrum in the burst frame ($E_{\rm p,z}$) into the $L_{\rm iso}-\Gamma_0$ relation, a tighter $L_{\rm iso}-E_{\rm p,z}-\Gamma_0$ relation is found (Liang et al. 2015). This relation places strong constraint on the composition of GRB fireballs.

The connection between long-duration GRBs (LGRBs) and SNe was predicted theoretically (Colgate 1974; Woosley 1993) and has been verified observationally (e.g., Galama et al. 1998; Hjorth et al. 2003; see review in Woosley \& Bloom   2006). They usually happen in the star formation regions of the galaxies (e.g., Paczy\'{n}ski 1998; see review in Woosley \& Bloom  2006 and Kumar \& Zhang 2015). The immensely bright afterglows illuminates the gas and dust within the star forming regions of the host galaxy and intervening intergalactic medium along the GRB line-of-sight. Their spectra are usually featureless power-law or broken power-law laws, which can be well described by the synchrotron radiations of relativistic electrons. Therefore, GRB afterglows are good probes of burst environment and the interstellar dust and gas in distant, star-forming galaxies (Jensen et al. 2001; Vreeswijk et al. 2004; Metzger et al. 1997; Savaglio et al. 2003; Chen, Hsiao-Wen et al. 2005; Schady et al. 2007, 2010; Starling et al. 2008; Prochaska et al. 2007a; Watson et al. 2007; Fox et al. 2008; Jang et al.2011; Xin et al., 2011). GRB afterglow spectra with Ly$\alpha$  absorption feature indicate the presence of large column densities of cold neutral
gas within GRB host galaxies, and their hydrogen column density ($N_{\rm H}$) are usually larger
than $10^{20.3}$ cm$^{-2}$ (e.g., Prochaska et al. 2007b; Schady et al. 2012). The damped Ly$\alpha$ systems (DLA) may represent the ISM near the GRBs in few kilo-parsecs(Kpc),
but not gas directly local to the GRB (Prochaska et al. 2007b). Thus GRB optical afterglows may be used as probes of the ISM in their host galaxies, as the ISM observed is less affected by the GRB or its progenitor (Watson et al. 2007). The visual dust extinctions ($A_{\rm V}$)
along the GRB lines of sight of many GRBs are low. As shown in Greiner et al. (2011), about 50\% GRBs
observed with GROND after the launch of {\em Swift} mission have $A_{\rm V}\sim 0$. In addition, the early optical light curves of about one-third GRBs show a smooth onset bump (Li et al. 2012). It may be due to the deceleration of the GRB fireball by the ambient medium (Sari \& Piran 1999; Kobayashi \& Zhang 2007). In this scenario, the rising slope of the bump is determined by the medium density profile ($n\propto r^{-k}$) and the spectrum index of the accelerated electrons [$N_e(E)\propto E^{-p}$], says, $\alpha=3-k(p+5)/4$ (Liang et al. 2013).  Hence, The afterglow onset bumps would be also an ideal probe to study the properties of the fireball and the profile of the circumburst medium. Liang et al. (2013) found that $k=0\sim 2$ (see also  Watson et al. 2007; Jin et al. 2012).

Although number of GRBs with very early X-ray and optical afterglow observations increases rapidly since the launch of the {\em Swift} mission, the sample is still rather smaller. GRBs with detection of very early afterglow, says, $\sim 100$ seconds and even earlier post the GRB trigger are valuable for revealing the fireball properties and its environment (Wang et al. 2013). This paper reports our very early optical observations of GRB 111228A (\S2). Using our observational data, together with the public prompt gamma-rays observed by both {\em Swift}/BAT and {\rm Fermi}/GBM, and the X-ray afterglows observed by {\em Swift}/XRT, we make data analysis and theoretical modeling to the optical and X-ray afterglows ($\S 3$) and analyze its properties of prompt gamma-rays (\S 4). Discussion is presented in \S 5, and conclusions are presented in \S6. Conventions $F\propto \nu^{\beta}t^{\alpha}$ and $Q_n=Q/10^{n}$ in cgs units are adopted.

\section{Optical Observations and Data Reduction}
GRB 111228A was detected by {\em Swift} Burst Alert Telescope (BAT) (Gehrels et al., 2004) at 15:44:43 UT on 2011 Dec. 28 (Ukwatta et al., 2011). It also triggered {\em Fermi} GBM at 15:45:30.80 UT (Briggs et al.,2011), 47.8 seconds  post the BAT trigger. Hereafter the zero time of our temporal analysis is set as the BAT trigger time. At 155 seconds after the trigger, {\em Swift}/UVOT  began to observe the burst and detected an optical counterpart whose coordinates are consistent with X-ray ones. We conducted a follow-up observation campaign of GRB 111228A with the TNT (0.8-m Tsinghua University - National Astronomical Observatory of China Telescope)at Xinglong Observatory\footnote{TNT is a 0.8-m telescope and runs by a custom-designed automation system for GRB follow-up observations (Zheng et al. 2008) at Xinglong Observatory. A PI $1300\times1340$ CCD and filters in the standard Johnson Bessel system are equipped for TNT.},  Lulin One-meter Telescope at Taiwan (Huang et al. 2005), and GMG 2.4m telescope at Lijiang observatory. We also made  ToO (Target of Opportunity) observations for this burst with CFHT telescope located at the summit of Mauna Kea of Hawaii.

Our earliest detection of the optical afterglow of GRB 111228A was obtained in the $W$ (white) and $R$ bands with TNT at 95 seconds post the {\em Swift}/BAT trigger. The first optical detection was during the last episode of the prompt gamma-rays. Its optical afterglows were also detected by LOT telescope  in four $SDSS$ filters, $g$, $r$, $i$ and $z$. We observed the afterglows of this burst with GMG 2.4m telescope  at later phase for three times, and several images in $B$,$V$ and $R$ band were obtained. Our ToO (Target of Opportunity) observations with
the CFHT telescope obtained the counterpart in the near-infrared wavelengths (the $J$ and $K$ bands ) at the later epoch.  To extend the temporal coverage of the optical afterglows late $R$-band data reported in GCN circular (Jang et al.,2011) were also collected  for our analysis.  The burst is at redshift of 0.714 (Dittmann et al. 2011; Cucchiara et al., 2011; Xu et al., 2011).

Our optical data reduction was carried out by following the standard routine in IRAF\footnote{IRAF is distributed by NOAO, which is operated by AURA, Inc., under cooperative agreement with NSF.} package, including bias and flat-field corrections. Point spread function (PSF) photometry was applied via the DAOPHOT task in the IRAF package to obtain the instrumental magnitudes. During the reduction, some frames were combined in order to increase the signal-to-noise ratio ($S/N$). In the calibration and analysis, the white band was treated as the $R$ band (Xin et al. 2010).
Absolute calibration for all data, excepted for $B,J,K$ bands data, was performed using the Sloan Digital Sky Survey ($SDSS$,
Adelman-McCarthy et al. 2008), with conversion of $SDSS$ to Johnson-Cousins
system\footnote{http://www.sdss.org/dr6/algorithms/sdssUBVRITransform.html
\#Lupton2005}. $J$ and $K$ bands data are calibrated by the nearby 2MASS catalog.
$B$-band data are calibrated by USNO B1.0 mag and a systematical error of 0.1 mag was adopted in our analysis.
The data of the optical observations are reported in Table 1.

\section{Properties of X-ray and optical Afterglows}
\subsection{Lightcurves}
GRB 111228A was observed by {\em Swift} XRT  since 145.1 sec after the GRB trigger, and its lightcurve and spectrum are extracted from the UK Swift Science Data Centre at the University of Leicester (Evans et al.
2009)\footnote{http://www.swift.ac.uk/results.shtml}. The optical and XRT lightcurves are  shown in Figure 1. The XRT lightcurve starts with an extremely rapid drop, as shown in the inset of Figure 1. It is most likely the tail emission of the prompt gamma-rays due to the arrival time delay of high latitude photons, the so called
curvature effect(e.g., Liang et al. 2006; Zhang et al. 2009). Following this segment an X-ray flare is observed, which could be further activity from the prompt emission. It also rapidly decays with a slope of $\alpha\sim-5.4$ after the peak of the flare\footnote{The slope is obtained by setting the zero time to the BAT trigger. It is physically meaningful only when we know the ejection time associated with the prompt pulse/flare, i.e., its intrinsic zero time (Liang et al. 2006).} No corresponding optical flare was simultaneously observed. The R-band optical lightcurve also starts with a steep decay ($\alpha\sim -5.4$ by setting the zero time to the BAT trigger time), but it is earlier than the initial steep decay in the X-ray lightcurve. The first optical data point was obtained during the last pulse of the prompt gamma-rays (from 95 to 115 seconds post the BAT trigger time).  Our joint spectral fit to the emission in both optical and BAT bands shows that it is probably prompt optical emission(see \S 4 for details). The optical flux increases, then transits to a long-lasting plateau.  The optical lightcurve is composed of multiple breaks (peaks), we empirically fit the lightcurve with a smooth broken power-law in two time intervals, i.e., $[150, 4500]$ and $[10^3, 5\times 10^5]$ seconds post the BAT trigger. The smooth broken power-law is taken as (e.g., Liang et al. 2007),
\begin{equation} F=F_0\left [
\left (   \frac{t}{t_b}\right)^{\omega\alpha_1}+\left (
\frac{t}{t_b}\right)^{\omega\alpha_2}\right]^{1/\omega}.
\end{equation}
We obtain $\alpha_{\rm O, 1}=2.62\pm 1.30$, $\alpha_{\rm O,2}=-0.17\pm 0.06$, $t_{\rm O, b, 1}=171\pm 9$ seconds, and $\omega_{\rm O,1}=3.08\pm 1.50$ with a reduced $\chi^2$ of 1.02 for the first time interval. The selected second time interval is partially overlapped with the first time interval. This is for fitting the data in the second time interval by fixing the $\alpha_{\rm O,1}$ as $\alpha_{\rm O,2}$. With this strategy, we then get $t_{\rm O, b, 2}=8527\pm 1392$ seconds, $\alpha_{\rm O,3}=-1.12\pm 0.10$, and $\omega_{\rm O,2}=1.59\pm 0.75$ with a reduced $\chi^2$ of 1.01. $t_{\rm O, b, 2}$ and $\omega_{\rm O,2}$ have large error bars. This is due to the broad, smooth transition from the shallowly-decaying phase to the normally-decaying phase. The X-ray lightcurve show similar features since $t>250$ seconds. It is well fitted with the smooth broken power-law, which yields $t_{\rm X, b}=8527$ s
(fixed\footnote{Being due to a large gap of the data in the X-ray plateau, the break time between the plateau and the
following normal decay segment is poorly constrained. We fix the break time the same as that in the optical band.}),
$\alpha_{\rm X, 1}=-0.24\pm 0.05$, $\alpha_{\rm X, 2}=-1.25\pm 0.03$, $\omega_{\rm X}=3.08\pm 1.50$ with a reduced $\chi^2=1.21$.

\subsection{Joint Optical-X-ray spectral Fits to the Afterglow data}
To investigate the spectral properties of the X-ray and optical afterglows, we extract the SEDs of the afterglows in the optical and X-ray band in four time intervals, i.e., 4-7 ks, 10-13 ks, 17-18.7 ks, and 62-88 ks. They are shown in Figure 2. The optical data are corrected for Galactic extinction with $E_{B-V}=0.032$ in the burst direction (Schlegel et al.1998).
We make joint spectral fit to the SEDs using the Xspec package. The extinction laws of the GRB host galaxy are taken as $R_V=3.16$ and $R_V=2.93$ for SMC and LMC, respectively. The hydrogen column density of Milky Way is fixed as $0.033\times10^{22}$cm$^{-2}$ along the line of sight to the burst.

The spectral model, chosen to fit the SEDs from optical to X-ray emission, is a single power-law or a broken power-law. Our results are reported in Table 2 and shown in Figure 2.
In the first two time intervals,
it is found that a broken power-law significantly improve the fits, comparing to the result by a single power-law. The indices are roughly 1.7 and 2.0 with a break at $\sim 1.0$ keV. The extinction of the host galaxy is almost negligible with $E(B-V)\sim 2 \times 10^{-2}$  for both LMC and SMC models. Note that in these spectra, a break at $\sim 1$ keV with small indices variation (from 1.7 to 2.0)  is likely due to being just a small curvature in the power law over several decades in frequency. In the last two episodes, a single power-law with photon index $\Gamma_{\rm OX}\sim 1.75$ is adequate to fit the SEDs. Therefore, we also show single power fits to the four SEDs in  Figure 2, with $\Gamma_{\rm OX}\sim 1.75$.

\subsection{Modeling the Afterglow Lightcurves}
As shown in Wang et al. (2015), the optical and X-ray lightcurves of most {\em Swift} GRBs are still consistent with the external shock models. In this section, we model the X-ray and optical lightcurves of GRB 111228A with the standard afterglow model by incorporating both the energy injection and jet opening anlgle effect. We assume that the GRB jet is observed along to the jet axis. The details of our model please refer to Sari et al. (1998) and Huang et al. (2000).   The Markov Chain Monte Carlo (MCMC) technique is used for our fitting. Our strategy is described as following.

\begin{itemize}
\item Deriving the index of the accelerated electron distribution from the observed spectral index and the slope of the normal decay phase of the lightcurves. As shown in Figure 1 and table 2, the optical and X-ray emission is in the same regime with a spectral index of $\beta\sim 0.75$, and the normal decay slope is $\alpha\sim 1.12$. It is found that they well satisfy the closure relation of $\alpha=3\beta/2$ in the standard fireball model, indicating that the emission is in the regime of $\nu_m<\nu<\nu_c$. Then, we have $p=2\beta+1=2.5$.

\item Explaining the observed shallow decay segment in both X-ray and optical lightcurves with a long-lasting energy injection. We describe the injection luminosity behavior as $L(t)=L_{\rm inj,0} t^{q}$ (e.g., Zhang \& M\'{e}sz\'{a}ros 2001). The observed decay slopes in the optical and X-ray bands are about $-0.15\sim -0.25$. We set $q\in(-0.15,-0.25)$ and $L_{\rm 0, inj}\in(10^{49},10^{51})$ erg/s in our fit. The break time between the shallow to the normal decay phases observed in the optical band have a large uncertainty and the transition is very smooth. Thus, we set the end time of the energy injection in a broad range, i.e., $t_{\rm end}\in (10^3, 10^4)$ seconds and assume that the injection is started at the BAT trigger time.
\item The other model parameters, i.e., $\Gamma_0$ (the initial Lorentz factor), $\epsilon_e$ (the faction of shock energy to electron), $\epsilon_B$ (the faction of shock energy to magnetic field), $n$ (the interstellar medium density), $\theta_j$ (jet break opening angle) are set in the ranges $\Gamma_0\in(100, 1000)$, $\epsilon_e\in(0.001, 0.5)$, $\epsilon_B\in(10^{-7},0.1)$, $n\in(0.001,10)$, $E_{\rm k}\in(10^{51}, 10^{53})$ erg, and $\theta_j\in (0.01,0.3)$ rad.
\item With the MCMC technique, we derive the best parameter set that can reproduce the lightcurves. We measure the goodness of our fits with a probability $p_{\rm f}=e^{-\chi^2/2}$, where $\chi^2$ is the reduced $\chi^2$. The time intervals of data for our fits are $t>700$ seconds for the X-ray data and $t>140$ seconds for the optical data.
\item Calculating the $1\sigma$ standard deviation of a given parameter by fixing the other parameters. The $1\sigma$ deviation then is derived from the Gaussian fit to the $p_f$ distribution.
\end{itemize}
Figure 3 shows the $p_f$ distributions along with our Gaussian fits for the best model parameters obtained with our MCMC technique. One can find that $p_f$ distributions of $\epsilon_e$,  $\epsilon_B$, $n$, $E_{\rm K}$, $q$, $L_{\rm 0, inj}$ are well fit with a Gaussian function. We have $\epsilon_e=(6.9\pm 0.3)\times 10^{-2}$, $\epsilon_B=(7.73\pm 0.62)\times 10^{-6}$, $E_{\rm K}=(6.32\pm 0.86)\times 10^{53}{\rm erg}$, $n=0.100\pm0.014$ $\rm cm^{-3}$,  $q=-0.18\pm 0.39$, $t_{\rm end}=5627^{+882}_{-688}$ s, and $\log L_{\rm 0, inj}=(4.66\pm 0.36)\times 10^{50}{\rm erg\ s^{-1}}$ . Note that $t_{\rm end}$ derived from our model fit is smaller than the break time from our empirical fit, i.e.,  $t_{\rm end}=5627^{+882}_{-688}$ seconds vs. $t_{\rm O, b, 2}=8527\pm 1392$ seconds. As we will discuss in Section 5, the energy injection may be contributed by the long-lasting injection from a newly-born pulsar. The dynamic evolution of the injected luminosity shows as an initial plateau that smoothly transits to a steep decay post the characteristic spin-down timescale (e.g., Dai \& Lu 1998; Zhang \& M{\'e}sz{\'a}ros 2001). This may result in a very smooth break in the afterglow observed lightcurves. The derived break time from our empirical fit thus is highly affected by the smooth parameter. $\Gamma_0$ and $\theta_j$ values mainly depend on the peak time of the early onset bump and late jet break time in both optical and X-ray afterglow data (e.g., Sari \& Piran 1999; Frail et al. 2001). The $p_f$ distribution of $\Gamma_0$ is  not a normal distribution. Its left and right sides can be separately fitted with a Gaussian function, which yields $\Gamma_0=476^{+225}_{-237}$. Our empirical fits to both the X-ray and optical lightcurves do not show any convincing jet break. Our model fit suggests $\theta_j\gtrsim 0.1$ rad. The above model parameters well fit the optical and X-ray lightcurves, with a reduce $\chi^2$ of 1.6\footnote{The large reduced $\chi^2$ is due to the fluctuation of early optical afterglow data in the time interval 140-320 seconds.}, as shown in Figure 4.

\section{Properties of prompt emission}
The BAT data of GRB 111228A are downloaded from the NASA {\em Swift} Achieve. We extract the BAT lightcurves with the standard {\em Swift} scientific tools. It is shown in Figure 5. We use the method of Bayesian Blocks (Scargle 1998) to analyze the temporal structure of the BAT lightcurve and identify three emission episodes. Its duration is $T_{90}=101.2\pm 5.42$ sec in the BAT band. No precursor or extended soft gamma-ray emission is observed.

Because of the narrowness of the BAT band, the BAT spectra are usually fit with a single power-law model only (e.g., Sakamoto et al. 2011; Zhang et al. 2007), we thus make spectral fit for the prompt gamma-rays with Fermi/GBM data. GBM comprises 12 NaI detectors covering an energy range from 8 keV to 1 MeV, and two BGO detectors sensitive to higher energies between 200 keV and 40 MeV. The GMB data are taken from the public  science support center at the NASA {\em Fermi data base} web site\footnote{ftp://legacy.gsfc.nasa.gov/fermi/data/}. The signals from all the 14 GBM detectors are collected by a central Data Processing Unit(DPU), which packages the resulting data into three different types: CTIME, CSPEC, and TTE. The CTIME accumulated spectral range is from 64 to 1024 ms, in multiples of 64 ms with 8 energy channels. The CSPEC data is binned in 1.024 s bins  with 128 energy channels. The TTE event data files contain individual photons with time and energy tags. We used the TTE data to make spectral fits by utilizing the software package RMFIT (version 3.3pr7). We made a joined spectral fit to the spectra collected by the NaI and BGO detectors with the Band function. User-defined intervals before and after the prompt emission phase were selected to obtain the background spectrum (e.g., Lyu et al. 2014).

GBM observed the 2nd and 3rd episodes of this GRB. The time-integrated spectrum observed with GBM and our fit with the Band function are shown in Figure 6. It is found that the Band function is adequate to fit the spectrum, with reduced $\chi^2=0.91$. The spectral parameters\footnote{$E_p$, $\beta_{\rm B,1}$ and $\beta_{\rm B,2}$ are the peak energy, low and high spectral index, respectively.} are $E_p=25.85\pm 2.97$ keV, $\beta_{\rm B,1}=-1.003\pm 0.327$, and $\beta_{\rm B,2}=-2.332\pm0.086$. We also fit the spectra accumulated in the 2nd and 3rd episodes, and derive $E_p=27.75\pm 3.66$ keV and $E_p=20.47\pm2.32$ keV, respectively.  The first optical data was detected during the 3rd episode. We make joint spectral fit for the spectrum in the optical and BAT bands (in the time interval from 95 to 115 second post the BAT trigger) with the Band function\footnote{We do not add the GBM data observed in the time interval in order to avoid possible calibration issue for different detectors.}. The Small Magellanic Cloud extinct law is adopted for measuring the host galaxy extinction in our fit. We find that a cutoff power-law model can well fit the broad band spectrum and the extinction of the host galaxy is negligible. We obtain a spectral index $\beta_{c}=-0.95\pm 0.02$ and a cutoff energy $E_c=18.4\pm 1.3$ keV, with a reduced $\chi^2$ of 1.38. The derived $E_p$ from the cutoff power-law model is $19.3\pm 1.4$ keV, which is consistent with that derived from the GBM data. This result likely indicates that the first optical data is the prompt emission.

The $E_p$ of the time-integrated spectrum of GRB111228A is $\sim 26$ keV, indicating that it is a soft GRB. Its peak luminosity in the $1-10^4$ keV band is $\log (L_{\rm iso,52}/{\rm erg\  s^{-1}})=-1.17\pm 0.02$. Liang et al. (2015) found a tight $L_{\rm iso}-E_{p,z}-\Gamma_0$ relation of long GRBs, i.e., $L^{\rm r}_{\rm iso, 52}=10^{-6.38\pm 0.35}{(E_{\rm p,z}/{\rm keV})}^{1.34\pm 0.14}\Gamma_{0}^{1.32\pm 0.19}$, where $E_{p, z}=E_p\times(1+z)$. We examine whether GRB 111228A satisfies this empirical relation. With $\log E_{p,z}=1.65\pm0.06$ and $\log \Gamma_0=2.68^{+0.21}_{-0.22}$, we have $\log (L^{\rm r}_{\rm iso, 52}/{\rm erg  s^{-1}})=-0.63_{-0.30}^{+0.29}$. As shown in Figure 7, it well satisfies the $L_{\rm iso}-E_{p,z}-\Gamma_0$ relation, and it bridges the typical GRBs and low luminosity GRBs in this relation.

\section{Discussion}
\subsection{Properties of the ambient medium}

Our spectral fits to the X-ray afterglows show $N_H$ of the host galaxy  along the line of sight is $(2.0\sim 2.5)\times 10^{21}$ cm$^{-2}$, which is higher than that of our Galaxy ($N_{\rm H}^{\rm WM}=3.3\times 10^{20}$ cm$^{-2}$) with one order of magnitude. However, the extinction of the host galaxy  along the line of sight is almost negligible ($E_{\rm B-V}=0.02\sim 0.06$), which is comparable to that of our Galaxy in the same direction. Therefore, the dust/gas ratio of the host galaxy of GRB 111228A in the line of sight is lower than that of our Galaxy one order of magnitude. Note the dust/gas ratio of the our Galaxy varies  in different line of sight, one still cannot convincingly suggest that the dust/gas ratio of the GRB host MUST be different from that of our Galaxy.  We also compare GRB 111228A with other GRBs in the $N_{\rm H}-A_{\rm V}$ plane in Figure 8, where the data of other GRBs are taken from Chen et al. (2014) and Greiner et al. (2011). It is found that GRB 111228A  does not show any distinct feature in comparison with the other typical GRBs in the $N_{\rm H}-A_{\rm V}$ plane.

The afterglow data indeed can be reproduced with the standard external shock models. The derived medium density is as low as $0.04$ cm$^{-3}$, likely indicating that the GRB is in a galactic halo or a hot interstellar medium. Panaitescu \& Kumar (2007) reported similar homogeneous media of low density for the afterglows of GRBs 990123 and 980703. Our analysis of the prompt gamma-rays of GRB 111228A suggests that it is a typical long GRBs. It is generally believed that the progenitor of long GRBs are massive stars (Woosley 1993; Paczy\'{n}ski
1998; MacFadyen \& Woosley 1999). In this model, the medium surrounding the bursts would be dense. This rises an issue that whether the low medium density observed in GRB 111228A still favors collapsar origin of this GRB? Scalo \& Wheeler (2001) proposed that the winds and supernovae are occurring in a cluster of massive stars may create 10 pc--1 kpc super bubbles. The density of the bubbles can be as low as
$10^{-3}$ cm$^{-3}$ and the circumburst medium
densities spanning a few orders of magnitude. Prochaska et al. (2007b) suggested that the GRB DLA may represent the ISM near the GRBs in few kpc,
but not gas directly local to the GRB. Modeling to the afterglow lightcurves of typical long GRBs show that the medium density significantly varies (e.g., Panaitescu \& Kumar 2001; Xin et al. 2012). Therefore, the derived low medium density may still favor the collapsar origin of GRB 111228A.

\subsection{Properties of the central engine}
A plateau phase is usually observed in the XRT lightcurves (Zhang et al. 2006; Liang et al. 2007) and in about one-third of optical lightcurves for long-duration GRBs (Li et al. 2012). Such a feature may be contributed to a long-lasting energy injection from a magnetar (Dai \& Lu 1998; Zhang \& M\'{e}sz\'{a}ros 2001). Observed late X-ray flares may also support this idea (Dai et al. 2006). L\"{u} \& Zhang (2014) systematically analyzed the
X-ray data and judged how likely a GRB might harbor a millisecond magnetar central engine. The achromatic optical and X-ray shallow decaying segments observed in GRB 111228A indicate that the central engine of this GRB may be a magnetar. With the observed spectral index and the decay slope of the plateau, we find that GRB 111228A is satisfied with criteria of ``Silver'' sample of magnetar central engine, following the method by L\"{u} \& Zhang (2014). The characteristic spin-down luminosity $L_0$ and spin-down time scale $\tau$ are related to the magnetar initial parameters as (Zhang \& M\'esz\'aros 2001)
\begin{equation}
 L_0 = 1.0 \times 10^{49}~{\rm erg~s^{-1}} (B_{p,15}^2 P_{0,-3}^{-4} R_6^6)
\label{L0}
\end{equation}
\begin{equation}
 \tau = 2.05 \times 10^3~{\rm s}~ (I_{45} B_{p,15}^{-2} P_{0,-3}^2 R_6^{-6})
\label{tau}
\end{equation}
where $B_p$ and $P_0$ are corresponding to surface polar cap magnetic field and initial spin period, respectively. From our lightcurve modeling, we take the end time of the energy injection phase as the spin-down time scale, i.e., $\tau=5267/(1+z)$ seconds, and take the spin-down luminosity as plateau luminosity as $\sim 4.66\times 10^{50}$ erg s$^{-1}$, and jet opening angle as 0.1 rad. We derive the $B_{\rm p} \sim 1.59\times 10^{15}$ G and $P_0 \sim 1.92$ ms. Figure 9 shows the comparison of GRB 111228A with a sample of GRBs available in  L\"{u} \& Zhang (2014) in the $B_p - P_0$ plane. It is found that GRB 111228A is a normal one with respect to the typical GRBs.

\section{Conclusion}
We have reported our optical observations of GRB 111228A from 95 sec to about 50 hours after the burst trigger. Using our data, together with prompt gamma-ray data observed by both {\em Swift}/BAT and {\rm Fermi}/GBM, and with the X-ray afterglows observed by {\em Swift}/XRT, we study the properties of the prompt gamma-rays and the ambient medium of GRB 111228A. We detected the prompt optical emission during the last episode of the burst. The optical transient rapidly faded, then gradually increased and peaked at around 200 seconds post the BAT trigger time. The optical lightcurve shows as a very smooth, slowly decaying bump without a sharp break. The XRT lightcurve is a canonic one, with a steep decay segment then transits to a very shall decay segment, being similar to the optical lightcurves since 250 second after the BAT trigger. The $E_p$ of the time-integrated spectrum of GRB111228A is $\sim 26$ keV, indicating that it is a soft GRB. It well satisfies the $L_{\rm iso}-E_{p,z}-\Gamma_0$ relation, and it bridges the typical GRBs and low luminosity GRBs in this relation. Our joint spectral fits to the X-ray and optical afterglow data show that the ambient medium features as low dust-to-gas ratio. Our theoretical modeling with the standard afterglow model by incorporating both the energy injection and jet break effect reveals that the medium density is $\sim$0.1 cm$^{-3}$,
which is much lower than that of typical long GRBs. The factions of the shock energy and electrons are comparable. Late energy injection with $L_{\rm in}\propto t^{-0.18\pm 0.39}$ up to $5627$ seconds is required to model the data, favoring a magnetar as the central engine of this GRB.

\section{Acknowledgement}
This work is supported by the National Basic Research Program of China (973 Program, grant No. 2014CB845800),
the National Natural Science Foundation of China (Grant No. 11533003, 11103036, U1331202, U1231115 and U1331101), the Strategic Priority Research Program ¡°The Emergence of Cosmological Structures¡± of the Chinese Academy of Sciences (grant XDB09000000), the Guangxi Science Foundation (Grant No. 2013GXNSFFA019001). The optical observations are partially supported by the Telescope Access
Program (TAP), which is funded by the National Astronomical Observatories, CAS, and the Special Fund for
Astronomy from the Ministry of Finance. We also acknowledge the use of the public data from the Swift data archive.
We appreciate helpful discussion with Xue-Feng Wu, Shu-Jin Hou, and Ji-Rong Mao.

\acknowledgments

\begin{deluxetable}{cccccc}
\tabletypesize{\small}
\tablewidth{0pt}
\label{Tab:pub-data}
\tablecaption{Optical Afterglow Photometry Log of GRB 090426,}
\tablehead{
\colhead{T-T0(mid,hour)} &
\colhead{Exposure (sec)} &
\colhead{Mag} &
\colhead{$\sigma$} &
\colhead{Filter} &
\colhead{Telescope}
}
\startdata
\object{0.0292}  &  20    &  16.297  &  0.081  &  W  &  TNT     \\
\object{0.0356}  &  20    &  17.115  &  0.149  &  W  &  TNT     \\
\object{0.0422}  &  20    &  17.285  &  0.145  &  W  &  TNT     \\
\object{0.0484}  &  20    &  16.980  &  0.119  &  W  &  TNT     \\
\object{0.0547}  &  20    &  16.995  &  0.112  &  W  &  TNT     \\
\object{0.0611}  &  20    &  16.930  &  0.095  &  W  &  TNT     \\
\object{0.0675}  &  20    &  16.904  &  0.123  &  W  &  TNT     \\
\object{0.0736}  &  20    &  16.817  &  0.115  &  W  &  TNT     \\
\object{0.0780}  &  20    &  16.897  &  0.103  &  W  &  TNT     \\
\object{0.0864}  &  20    &  16.805  &  0.091  &  W  &  TNT     \\
\object{0.0925}  &  20    &  16.907  &  0.128  &  W  &  TNT     \\
\object{0.0989}  &  20    &  17.191  &  0.145  &  W  &  TNT     \\
\object{0.1053}   &  20    &  16.931  &  0.120  &  W  &  TNT     \\
\object{0.1117}   &  20    &  17.085  &  0.150  &  W  &  TNT     \\
\object{0.1181}   &  20    &  17.084  &  0.111  &  W  &  TNT     \\
\object{0.1242}   &  20    &  17.056  &  0.149  &  W  &  TNT     \\
\object{0.1306}   &  20    &  16.930  &  0.132  &  W  &  TNT     \\
\object{0.1370}   &  20    &  16.969  &  0.134  &  W  &  TNT     \\
\object{0.1434}   &  20    &  16.810  &  0.086  &  W  &  TNT     \\
\object{0.1495}   &  20    &  17.224  &  0.146  &  W  &  TNT     \\
\object{0.1636}   &  60    &  17.055  &  0.081  &  R  &  TNT     \\
\object{0.1855}   &  60    &  17.015  &  0.095  &  R  &  TNT     \\
\object{0.2075}   &  60    &  16.996  &  0.069  &  R  &  TNT     \\
\object{0.2294}   &  60    &  17.096  &  0.113  &  R  &  TNT     \\
\object{0.2511}   &  60    &  17.224  &  0.095  &  R  &  TNT     \\
\object{0.2730}   &  60    &  17.180  &  0.091  &  R  &  TNT     \\
\object{0.2947}   &  60    &  17.218  &  0.104  &  R  &  TNT     \\
\object{0.3164}   &  60    &  17.160  &  0.110  &  R  &  TNT     \\
\object{0.3383}   &  60    &  17.231  &  0.107  &  R  &  TNT     \\
\object{0.3602}   &  60    &  17.117  &  0.080  &  R  &  TNT     \\
\object{0.3819}   &  60    &  17.295  &  0.123  &  R  &  TNT     \\
\object{0.4036}   &  60    &  17.375  &  0.125  &  R  &  TNT     \\
\object{0.4255}   &  60    &  17.233  &  0.137  &  R  &  TNT     \\
\object{0.4472}   &  60    &  17.271  &  0.130  &  R  &  TNT     \\
\object{0.4691}   &  60    &  17.425  &  0.157  &  R  &  TNT     \\
\object{0.4908}   &  60    &  17.500  &  0.149  &  R  &  TNT     \\
\object{0.5127}   &  60    &  17.200  &  0.129  &  R  &  TNT     \\
\object{0.5344}   &  60    &  17.395  &  0.141  &  R  &  TNT     \\
\object{0.5564}   &  60    &  17.153  &  0.107  &  R  &  TNT     \\
\object{0.5780}   &  60    &  17.337  &  0.134  &  R  &  TNT     \\
\object{0.1636}   &  60    &  17.055  &  0.081  &  R  &  TNT     \\
\object{0.1855}   &  60    &  17.015  &  0.095  &  R  &  TNT     \\
\object{0.2075}   &  60    &  16.996  &  0.069  &  R  &  TNT     \\
\object{0.2294}   &  60    &  17.096  &  0.113  &  R  &  TNT     \\
\object{0.2511}   &  60    &  17.224  &  0.095  &  R  &  TNT     \\
\object{0.2730}   &  60    &  17.180  &  0.091  &  R  &  TNT     \\
\object{0.2947}   &  60    &  17.218  &  0.104  &  R  &  TNT     \\
\object{0.3164}   &  60    &  17.160  &  0.110  &  R  &  TNT     \\
\object{0.3383}   &  60    &  17.231  &  0.107  &  R  &  TNT     \\
\object{0.3602}   &  60    &  17.117  &  0.080  &  R  &  TNT     \\
\object{0.3819}   &  60    &  17.295  &  0.123  &  R  &  TNT     \\
\object{0.4036}   &  60    &  17.375  &  0.125  &  R  &  TNT     \\
\object{0.4255}   &  60    &  17.233  &  0.137  &  R  &  TNT     \\
\object{0.4472}   &  60    &  17.271  &  0.130  &  R  &  TNT     \\
\object{0.4691}   &  60    &  17.425  &  0.157  &  R  &  TNT     \\
\object{0.4908}   &  60    &  17.500  &  0.149  &  R  &  TNT     \\
\object{0.5127}   &  60    &  17.200  &  0.129  &  R  &  TNT     \\
\object{0.5344}   &  60    &  17.395  &  0.141  &  R  &  TNT     \\
\object{0.5564}   &  60    &  17.153  &  0.107  &  R  &  TNT     \\
\object{0.5780}   &  60    &  17.337  &  0.134  &  R  &  TNT     \\
\object{0.6336}   &  300   &  17.272  &  0.056  &  R  &  TNT     \\
\object{0.7220}   &  300   &  17.460  &  0.056  &  R  &  TNT     \\
\object{0.9120}   &  300   &  17.549  &  0.059  &  R  &  TNT     \\
\object{1.0902}    &  300   &  17.550  &  0.058  &  R  &  TNT     \\
\object{1.2686}    &  300   &  17.603  &  0.060  &  R  &  TNT     \\
\object{1.4467}    &  300   &  17.817  &  0.066  &  R  &  TNT     \\
\object{1.6250}    &  300   &  17.813  &  0.060  &  R  &  TNT     \\
\object{1.8034}    &  300   &  17.860  &  0.058  &  R  &  TNT     \\
\object{1.9817}    &  300   &  17.943  &  0.058  &  R  &  TNT     \\
\object{2.1598}    &  300   &  17.981  &  0.073  &  R  &  TNT     \\
\object{2.3381}    &  300   &  18.097  &  0.108  &  R  &  TNT     \\
\object{2.5164}    &  300   &  18.203  &  0.077  &  R  &  TNT     \\
\object{2.6945}    &  300   &  18.120  &  0.071  &  R  &  TNT     \\
\object{2.8728}    &  300   &  18.153  &  0.067  &  R  &  TNT     \\
\object{3.0511}    &  300   &  18.128  &  0.057  &  R  &  TNT     \\
\object{3.2295}    &  300   &  18.222  &  0.062  &  R  &  TNT     \\
\object{3.4075}    &  300   &  18.309  &  0.073  &  R  &  TNT     \\
\object{3.5859}    &  300   &  18.406  &  0.085  &  R  &  TNT     \\
\object{3.7642}    &  300   &  18.297  &  0.118  &  R  &  TNT     \\
\object{3.9423}    &  300   &  18.163  &  0.142  &  R  &  TNT     \\
\object{4.2989}    &  300   &  18.386  &  0.132  &  R  &  TNT     \\
\object{4.4773}    &  300   &  18.489  &  0.139  &  R  &  TNT     \\
\object{4.6553}    &  300   &  18.382  &  0.098  &  R  &  TNT     \\
\object{4.8336}    &  300   &  18.705  &  0.124  &  R  &  TNT     \\
\object{5.0120}    &  300   &  18.487  &  0.514  &  R  &  TNT     \\
\object{5.1900}    &  300   &  18.428  &  0.142  &  R  &  TNT     \\
\object{5.9031}    &  300   &  18.756  &  0.098  &  R  &  TNT     \\
\object{0.7792}   &  300   &  17.811   &  0.080  &  V  &  TNT\\
\object{0.9594}   &  300   &  17.724   &  0.074  &  V  &  TNT\\
\object{1.1378}   &  300   &  18.008   &  0.087  &  V  &  TNT\\
\object{1.3161}   &  300   &  18.031   &  0.094  &  V  &  TNT\\
\object{1.4942}   &  300   &  18.017   &  0.087  &  V  &  TNT\\
\object{1.6725}   &  300   &  18.188   &  0.074  &  V  &  TNT\\
\object{1.8508}   &  300   &  18.183   &  0.083  &  V  &  TNT\\
\object{2.0289}   &  300   &  18.356   &  0.074  &  V  &  TNT\\
\object{2.2072}   &  300   &  18.450   &  0.107  &  V  &  TNT\\
\object{2.3856}   &  300   &  18.542   &  0.123  &  V  &  TNT\\
\object{2.5639}   &  300   &  18.618   &  0.126  &  V  &  TNT\\
\object{2.7419}   &  300   &  18.574   &  0.101  &  V  &  TNT\\
\object{2.9203}   &  300   &  18.406   &  0.056  &  V  &  TNT\\
\object{3.0986}   &  300   &  18.508   &  0.065  &  V  &  TNT\\
\object{3.2767}   &  300   &  18.581   &  0.069  &  V  &  TNT\\
\object{3.4550}   &  300   &  18.886   &  0.133  &  V  &  TNT\\
\object{3.6333}   &  300   &  18.574   &  0.086  &  V  &  TNT\\
\object{3.8117}   &  300   &  18.559   &  0.185  &  V  &  TNT\\
\object{4.3464}   &  300   &  18.838   &  0.187  &  V  &  TNT\\
\object{4.5244}   &  300   &  18.938   &  0.129  &  V  &  TNT\\
\object{4.7028}   &  300   &  18.733   &  0.125  &  V  &  TNT\\
\object{4.8811}   &  300   &  19.104   &  0.151  &  V  &  TNT\\
\object{5.5942}   &  300   &  18.982   &  0.107  &  V  &  TNT\\
\object{5.7722}   &  300   &  19.499   &  0.236  &  V  &  TNT\\
\object{5.9506}   &  300   &  19.143   &  0.123  &  V  &  TNT\\
\object{24.8198}  &  600   &  20.946  &  0.060   &  B  &  LJ2.4m\\
\object{25.4064}  &  600   &  21.064  &  0.064   &  B  &  LJ2.4m\\
\object{26.0269}  &  600   &  21.076  &  0.062   &  B  &  LJ2.4m\\
\object{50.7076}  &  1800  &  22.026  &  0.100   &  B  &  LJ2.4m\\
\object{24.6368}  &  600   &  20.600   &  0.060   &  V  &  LJ2.4m\\
\object{25.2281}  &  600   &  20.755   &  0.063   &  V  &  LJ2.4m\\
\object{25.8323}  &  600   &  20.715   &  0.061   &  V  &  LJ2.4m\\
\object{50.5283}  &  1800  &  21.615   &  0.104   &  V  &  LJ2.4m\\
\object{0.8654}   &  600   &  17.440   &  0.010   &  R  &  LJ2.4m  \\
\object{1.5753}    &  600   &  17.742   &  0.01   &  R  &  LJ2.4m  \\
\object{2.0040}    &  600   &  17.885  &  0.01   &  R  &  LJ2.4m  \\
\object{2.1923}    &  600   &  17.950   &  0.01   &  R  &  LJ2.4m  \\
\object{24.4650}     &  600   &  20.431   &  0.06   &  R  &  LJ2.4m  \\
\object{25.1342}    &  600   &  20.383   &  0.06   &  R  &  LJ2.4m  \\
\object{25.7355}    &  600   &  20.532   &  0.06   &  R  &  LJ2.4m  \\
\object{50.5942}    &  1800  &  21.100    &  0.10   &  R  &  LJ2.4m  \\
\object{0.311}  &  180.00  &  17.700  &  0.019  &  SDSS-g  &  LOT  \\
\object{0.505}  &  180.00  &  17.782  &  0.021  &  SDSS-g  &  LOT  \\
\object{0.697}  &  180.00  &  17.859  &  0.020  &  SDSS-g  &  LOT  \\
\object{0.893}  &  180.00  &  17.913  &  0.021  &  SDSS-g  &  LOT  \\
\object{1.089}  &  180.00  &  18.031  &  0.021  &  SDSS-g  &  LOT  \\
\object{1.350}  &  180.00  &  18.139  &  0.018  &  SDSS-g  &  LOT  \\
\object{1.647}  &  180.00  &  18.278  &  0.017  &  SDSS-g  &  LOT  \\
\object{1.946}  &  180.00  &  18.369  &  0.017  &  SDSS-g  &  LOT  \\
\object{2.239}  &  180.00  &  18.453  &  0.019  &  SDSS-g  &  LOT  \\
\object{2.538}  &  180.00  &  18.557  &  0.018  &  SDSS-g  &  LOT  \\
\object{2.834}  &  180.00  &  18.650  &  0.018  &  SDSS-g  &  LOT  \\
\object{3.134}  &  180.00  &  18.740  &  0.018  &  SDSS-g  &  LOT  \\
\object{3.481}  &  180.00  &  18.808  &  0.021  &  SDSS-g  &  LOT  \\
\object{3.779}  &  180.00  &  18.880  &  0.021  &  SDSS-g  &  LOT  \\
\object{4.079}  &  180.00  &  18.973  &  0.020  &  SDSS-g  &  LOT  \\
\object{4.375}  &  180.00  &  19.055  &  0.019  &  SDSS-g  &  LOT  \\
\object{4.671}  &  180.00  &  19.086  &  0.021  &  SDSS-g  &  LOT  \\
\object{4.968}  &  180.00  &  19.132  &  0.020  &  SDSS-g  &  LOT  \\
\object{5.266}  &  180.00  &  19.171  &  0.024  &  SDSS-g  &  LOT  \\
\object{0.376}  &  180.00  &  17.381  &  0.019  &  SDSS-r  &  LOT  \\
\object{0.569}  &  180.00  &  17.450  &  0.018  &  SDSS-r  &  LOT  \\
\object{0.763}  &  180.00  &  17.509  &  0.019  &  SDSS-r  &  LOT  \\
\object{0.958}  &  180.00  &  17.624  &  0.020  &  SDSS-r  &  LOT  \\
\object{1.153}  &  180.00  &  17.701  &  0.021  &  SDSS-r  &  LOT  \\
\object{1.414}  &  180.00  &  17.787  &  0.017  &  SDSS-r  &  LOT  \\
\object{1.713}  &  180.00  &  17.938  &  0.018  &  SDSS-r  &  LOT  \\
\object{2.011}  &  180.00  &  18.024  &  0.018  &  SDSS-r  &  LOT  \\
\object{2.306}  &  180.00  &  18.126  &  0.019  &  SDSS-r  &  LOT  \\
\object{2.601}  &  180.00  &  18.206  &  0.020  &  SDSS-r  &  LOT  \\
\object{2.899}  &  180.00  &  18.270  &  0.019  &  SDSS-r  &  LOT  \\
\object{3.199}  &  180.00  &  18.332  &  0.022  &  SDSS-r  &  LOT  \\
\object{3.547}  &  180.00  &  18.430  &  0.019  &  SDSS-r  &  LOT  \\
\object{3.845}  &  180.00  &  18.495  &  0.020  &  SDSS-r  &  LOT  \\
\object{4.144}  &  180.00  &  18.562  &  0.021  &  SDSS-r  &  LOT  \\
\object{4.439}  &  180.00  &  18.643  &  0.023  &  SDSS-r  &  LOT  \\
\object{4.734}  &  180.00  &  18.684  &  0.024  &  SDSS-r  &  LOT  \\
\object{5.032}  &  180.00  &  18.816  &  0.026  &  SDSS-r  &  LOT  \\
\object{5.332}  &  180.00  &  18.802  &  0.025  &  SDSS-r  &  LOT  \\
\object{0.439}  &  180.00  &  17.173  &  0.025  &  SDSS-i  &  LOT  \\
\object{0.634}  &  180.00  &  17.301  &  0.023  &  SDSS-i  &  LOT  \\
\object{0.827}  &  180.00  &  17.341  &  0.025  &  SDSS-i  &  LOT  \\
\object{1.023}  &  180.00  &  17.523  &  0.027  &  SDSS-i  &  LOT  \\
\object{1.216}  &  180.00  &  17.555  &  0.028  &  SDSS-i  &  LOT  \\
\object{1.479}  &  180.00  &  17.667  &  0.021  &  SDSS-i  &  LOT  \\
\object{1.778}  &  180.00  &  17.786  &  0.021  &  SDSS-i  &  LOT  \\
\object{2.074}  &  180.00  &  17.868  &  0.023  &  SDSS-i  &  LOT  \\
\object{2.371}  &  180.00  &  17.963  &  0.024  &  SDSS-i  &  LOT  \\
\object{2.663}  &  180.00  &  18.011  &  0.023  &  SDSS-i  &  LOT  \\
\object{2.965}  &  180.00  &  18.134  &  0.027  &  SDSS-i  &  LOT  \\
\object{3.263}  &  180.00  &  18.192  &  0.028  &  SDSS-i  &  LOT  \\
\object{3.610}  &  180.00  &  18.247  &  0.024  &  SDSS-i  &  LOT  \\
\object{3.908}  &  180.00  &  18.371  &  0.026  &  SDSS-i  &  LOT  \\
\object{4.208}  &  180.00  &  18.446  &  0.026  &  SDSS-i  &  LOT  \\
\object{4.504}  &  180.00  &  18.396  &  0.029  &  SDSS-i  &  LOT  \\
\object{4.798}  &  180.00  &  18.512  &  0.028  &  SDSS-i  &  LOT  \\
\object{5.097}  &  180.00  &  18.549  &  0.029  &  SDSS-i  &  LOT  \\
\object{5.396}  &  180.00  &  18.636  &  0.034  &  SDSS-i  &  LOT  \\
\object{1.565}  &  300.00  &  17.457  &  0.030  &  SDSS-z  &  LOT  \\
\object{1.863}  &  300.00  &  17.589  &  0.028  &  SDSS-z  &  LOT  \\
\object{2.157}  &  300.00  &  17.570  &  0.029  &  SDSS-z  &  LOT  \\
\object{2.456}  &  300.00  &  17.763  &  0.031  &  SDSS-z  &  LOT  \\
\object{2.749}  &  300.00  &  17.858  &  0.033  &  SDSS-z  &  LOT  \\
\object{3.050}  &  300.00  &  17.937  &  0.034  &  SDSS-z  &  LOT  \\
\object{3.348}  &  300.00  &  17.969  &  0.039  &  SDSS-z  &  LOT  \\
\object{3.695}  &  300.00  &  18.057  &  0.045  &  SDSS-z  &  LOT  \\
\object{3.995}  &  300.00  &  18.250  &  0.042  &  SDSS-z  &  LOT  \\
\object{4.294}  &  300.00  &  18.237  &  0.040  &  SDSS-z  &  LOT  \\
\object{4.587}  &  300.00  &  18.458  &  0.044  &  SDSS-z  &  LOT  \\
\object{4.883}  &  300.00  &  18.356  &  0.040  &  SDSS-z  &  LOT  \\
\object{5.182}  &  300.00  &  18.462  &  0.046  &  SDSS-z  &  LOT  \\
\object{5.483}  &  300.00  &  18.469  &  0.044  &  SDSS-z  &  LOT  \\
\object{19.5149}  &  29.5  &  18.595  &  0.030             &  J  &  CFHT  \\
\object{19.6680}   &  29.5  &  18.675  &  0.028            &  J  &  CFHT  \\
\object{19.7438}  &  29.5  &  18.642  &  0.025            &  J  &  CFHT  \\
\object{19.8216}  &  29.5  &  18.654  &  0.028            &  J  &  CFHT  \\
\object{19.9781}  &  29.5  &  18.674  &  0.027            &  J  &  CFHT  \\
\object{20.0539}  &  29.5  &  18.692  &  0.027            &  J  &  CFHT  \\
\object{20.2066}  &  29.5  &  18.692  &  0.029            &  J  &  CFHT  \\
\object{20.2834}  &  29.5  &  18.670   &  0.029            &  J  &  CFHT  \\
\object{20.3609}  &  29.5  &  18.659  &  0.026            &  J  &  CFHT  \\
\object{20.5133}  &  29.5  &  18.724  &  0.025            &  J  &  CFHT  \\
\object{20.5908}  &  29.5  &  18.741  &  0.025            &  J  &  CFHT  \\
\object{20.7425}  &  29.5  &  18.814  &  0.029            &  J  &  CFHT  \\
\object{20.8010}   &  29.5  &  17.213  &  0.029            &  K  &  CFHT  \\
\object{20.8582}  &  29.5  &  17.223  &  0.026            &  K  &  CFHT  \\
\object{20.9136}  &  29.5  &  17.250   &  0.027            &  K  &  CFHT  \\
\object{20.9707}  &  29.5  &  17.212  &  0.028            &  K  &  CFHT  \\
\object{21.0262}  &  29.5  &  17.265  &  0.026            &  K  &  CFHT  \\
\object{21.0804}  &  29.5  &  17.265  &  0.025            &  K  &  CFHT  \\
\object{21.1378}  &  29.5  &  17.257  &  0.024            &  K  &  CFHT  \\
\object{21.1934}  &  29.5  &  17.272  &  0.026            &  K  &  CFHT  \\
\object{21.2489}  &  29.5  &  17.280   &  0.025            &  K  &  CFHT  \\
\object{21.3046}  &  29.5  &  17.298  &  0.024            &  K  &  CFHT  \\
\object{21.3614}  &  29.5  &  17.229  &  0.024            &  K  &  CFHT  \\
\object{21.4169}  &  29.5  &  17.295  &  0.026            &  K  &  CFHT  \\
\object{21.4730}   &  29.5  &  17.260   &  0.027            &  K  &  CFHT  \\
\object{21.5304}  &  29.5  &  17.271  &  0.025            &  K  &  CFHT  \\
\object{21.6986}  &  29.5  &  17.306  &  0.024            &  K  &  CFHT  \\
\object{21.7541}  &  29.5  &  17.270   &  0.024            &  K  &  CFHT  \\
\object{21.8110}   &  29.5  &  17.278  &  0.023            &  K  &  CFHT  \\
\object{21.8662}  &  29.5  &  17.328  &  0.025            &  K  &  CFHT  \\
\enddata
\tablecomments{
The reference time $T_0$ is {\em Swift} BAT burst trigger time. "T-T0" is the middle time in hour for each data. "Exposure" is the exposure time for each data in second. All Data are not corrected for the Galactic extinction (which is $E_{B-V}=0.03$, Schlegel et al.1998).
B-band data is calibrated by USNO B1.0 B2mag with 2 stars. Other bands data is  calibrated by SDSS reference stars. The magnitudes in SDSS filters are in AB magnitude system, and others are in the Vega magnitude system.}
\end{deluxetable}

\begin{deluxetable}{ccccccc}
\tabletypesize{\footnotesize}
\tablewidth{500pt}
\label{Tab:pub-data}
\tablecaption{Fitting Result of Optical and X-ray SEDs}
\tablehead{
\colhead{Interval(s)} &
\colhead{Model($\chi^{2}/{\rm dof}$)} &
\colhead{E(B-V)($10^{-2}$)} &
\colhead{$N_{\rm H}^{\rm host}$($10^{22}{\rm cm}^{-2}$)} &
\colhead{$\Gamma_{\rm OX,1}$} &
\colhead{$E_{b}(\rm KeV)$} &
\colhead{$\Gamma_{\rm OX,2}$}
}
\startdata
\object{4k-7k}  &  Lmc*Bkn($105.88/33$)    &  $2.19_{-0.39}^{+0.39}$  &  0.225  &  $1.69_{-0.003}^{+0.003}$  &  $1.05_{-0.17}^{+0.17}$  &  2.01     \\
                &  Smc*Bkn($106.56/33$)    &  $2.32_{-0.42}^{+0.42}$  &  0.225  &  $1.69_{-0.003}^{+0.003}$  &  $1.05_{-0.17}^{+0.17}$  &  2.01     \\
\object{10k-13k}  &  Lmc*Bkn($97.28/58$)    &  $6.12_{-0.70}^{+0.71}$  &  0.242  &  $1.67_{-0.006}^{+0.006}$  &  $1.04_{-0.18}^{+0.24}$  &  2.00     \\
                  &  Smc*Bkn($96.25/58$)    &  $6.58_{-0.75}^{+0.75}$  &  0.242  &  $1.67_{-0.006}^{+0.006}$  &  $1.04_{-0.18}^{+0.24}$  &  2.00     \\
\object{17k-18.7k}  &  Lmc*PL($19.74/29$)    &  $4.94_{-1.20}^{+1.21}$  &  $0.213_{-0.088}^{+0.122}$  &  $--$  &  $--$  &  $1.76_{-0.018}^{+0.019}$     \\
                  &  Smc*PL($20.31/29$)    &  $5.26_{-1.27}^{+1.29}$  &  $0.213_{-0.088}^{+0.122}$  &  $--$  &  $--$  &  $1.76_{-0.018}^{+0.019}$     \\
\object{62k-88k}  &  PL($63.72/41$)    &  $--$  &  $0.254_{-0.083}^{+0.112}$  &  $--$  &  $--$  &  $1.74_{-0.010}^{+0.011}$     \\

\enddata
\end{deluxetable}


%

\clearpage

\begin{figure}[htbp]
 \centering
\includegraphics[angle=0,width=0.8\textwidth]{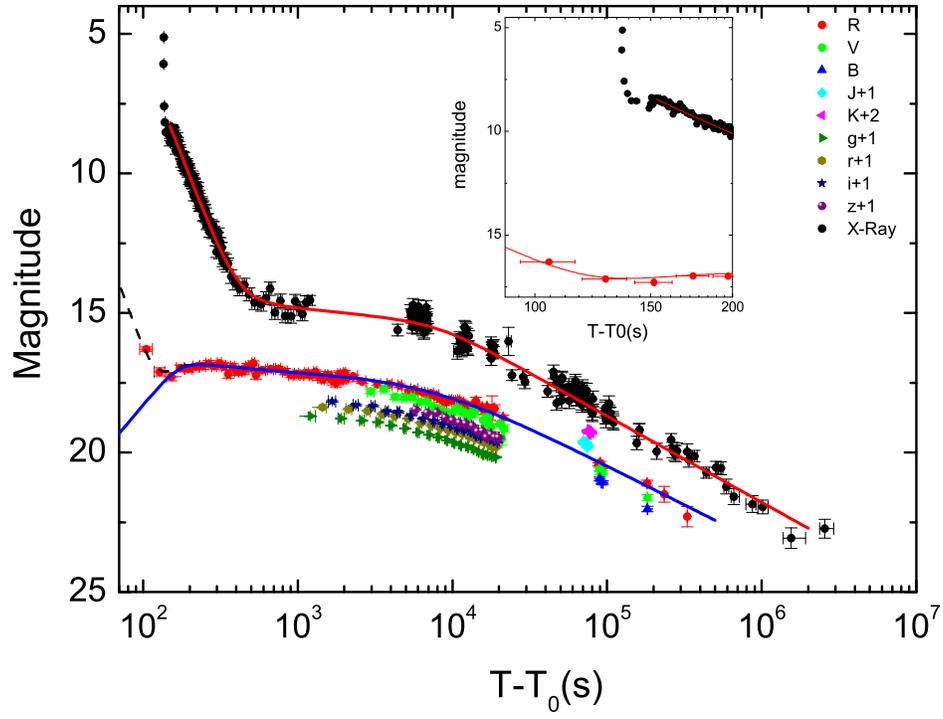}
\caption{X-ray and optical afterglow light curves of GRB 111228A. The solid lines are our empirical fits with power-law functions. The {\em inset} is first 200 seconds of X-ray and optical lightcurves.}
\end{figure}

 \begin{figure}[htbp]
 \centering
\includegraphics[angle=0,width=0.8\textwidth]{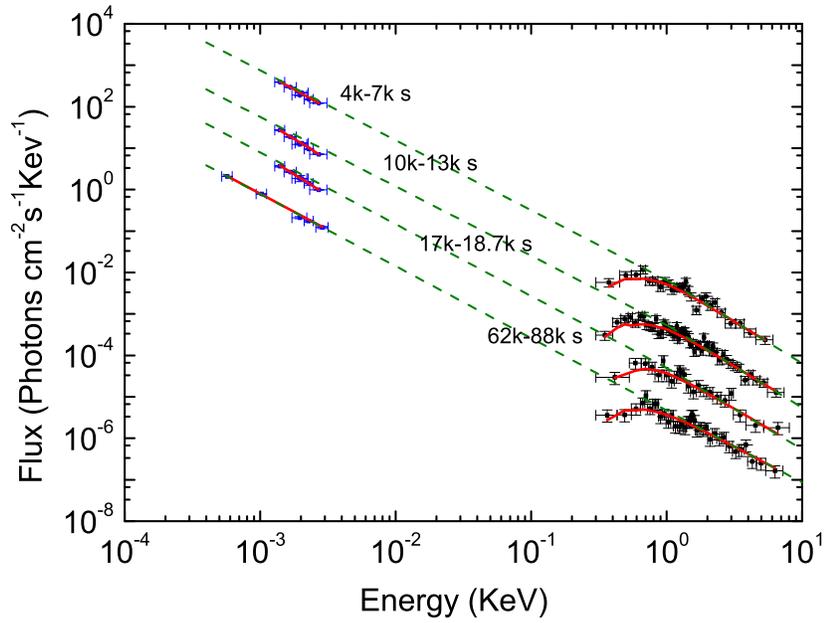}
\caption{Spectral energy distributions of the optical-X-ray afterglows in four selected
time intervals and our joint spectral fit with a single power-law or a broken power-law models ({\em solid lines}).
Dashed lines are fits with absorbed power-law function that are extrapolated to the optical bands.\label{Opt_X_SEDs}}
\end{figure}%

 \begin{figure}[htbp]
 \centering
\includegraphics[angle=0,width=0.3\textwidth]{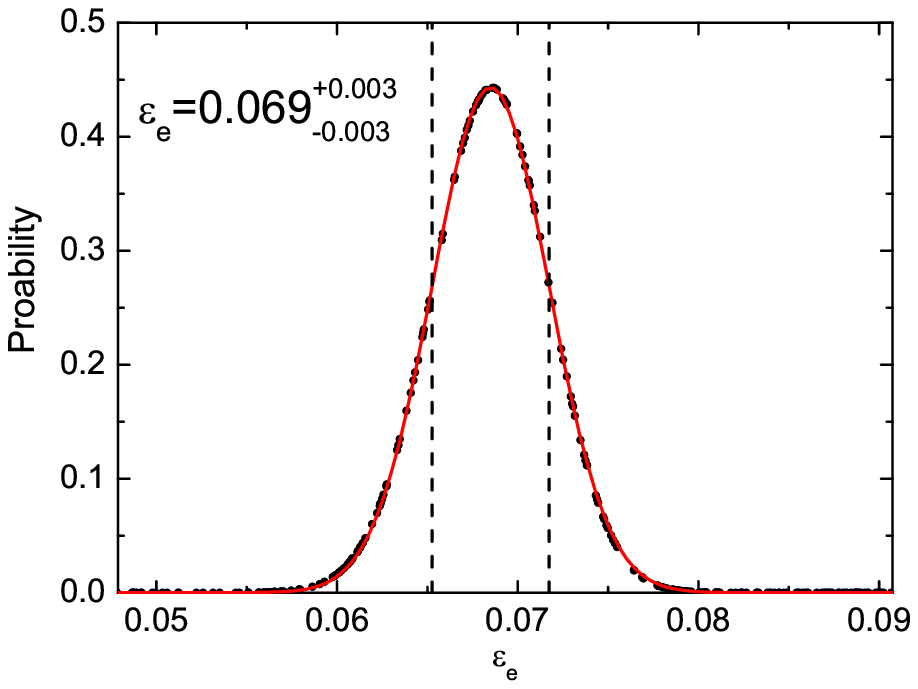}
\includegraphics[angle=0,width=0.3\textwidth]{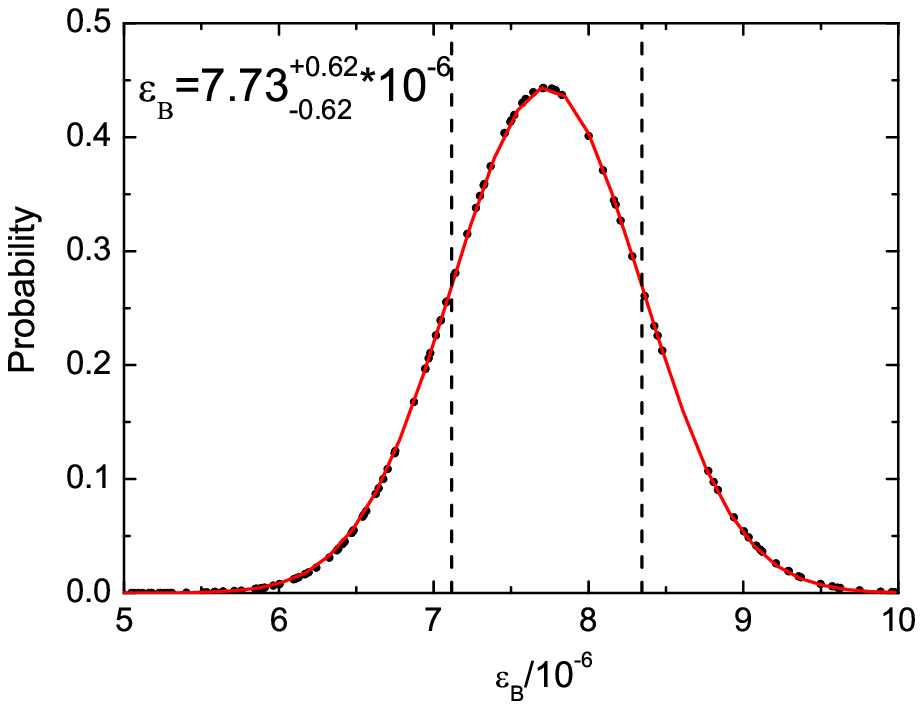}
\includegraphics[angle=0,width=0.3\textwidth]{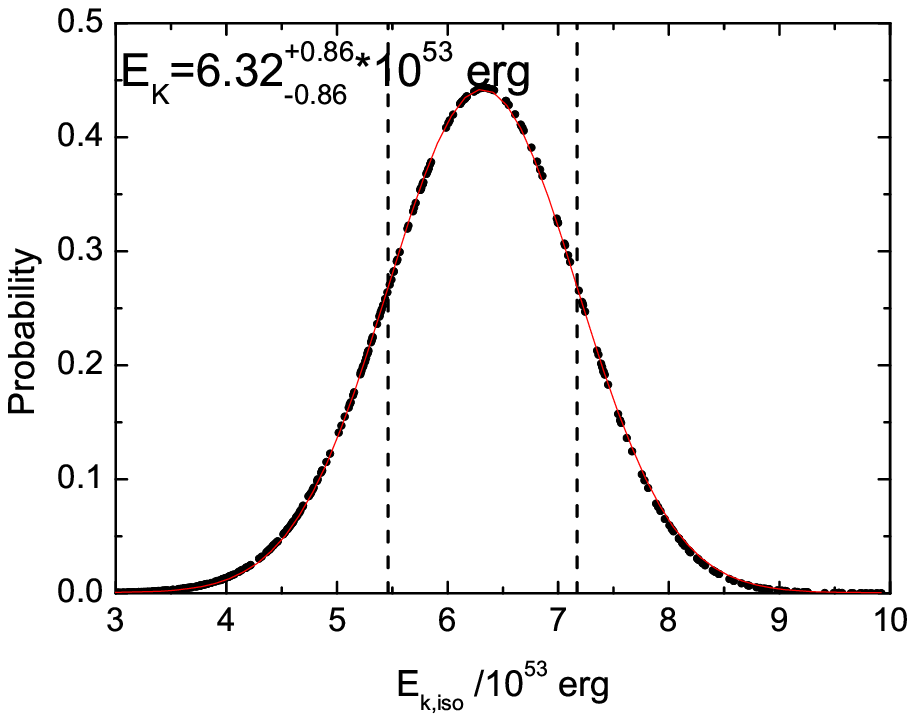}
\includegraphics[angle=0,width=0.3\textwidth]{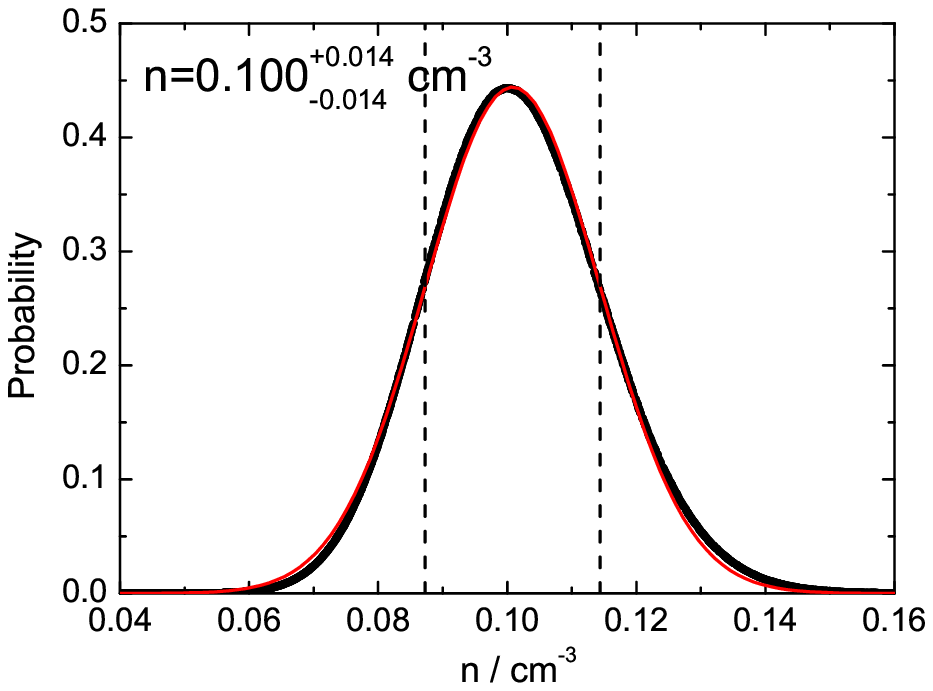}
\includegraphics[angle=0,width=0.3\textwidth]{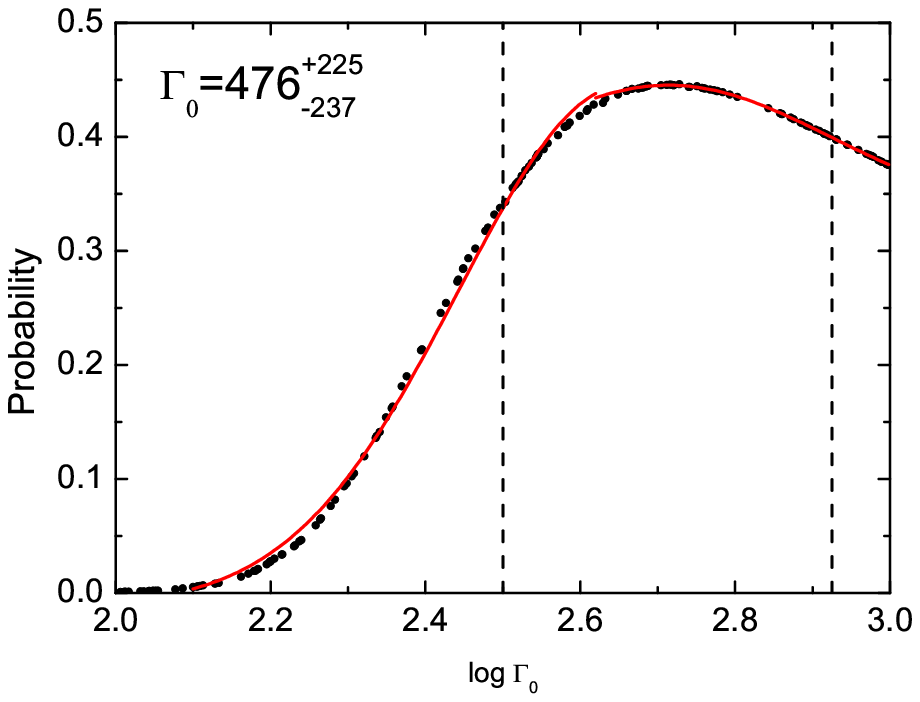}
\includegraphics[angle=0,width=0.3\textwidth]{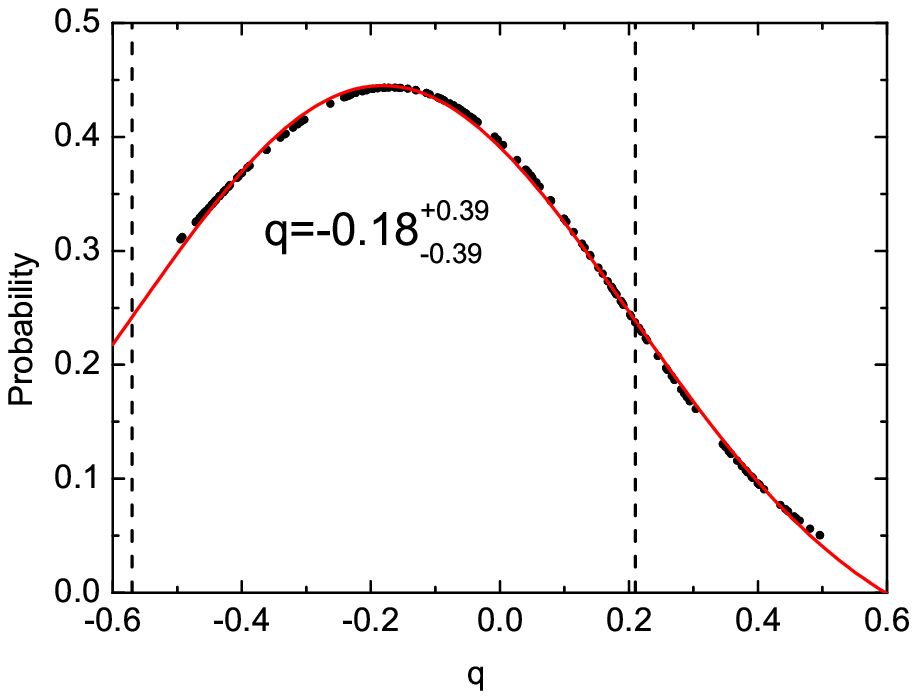}
\includegraphics[angle=0,width=0.3\textwidth]{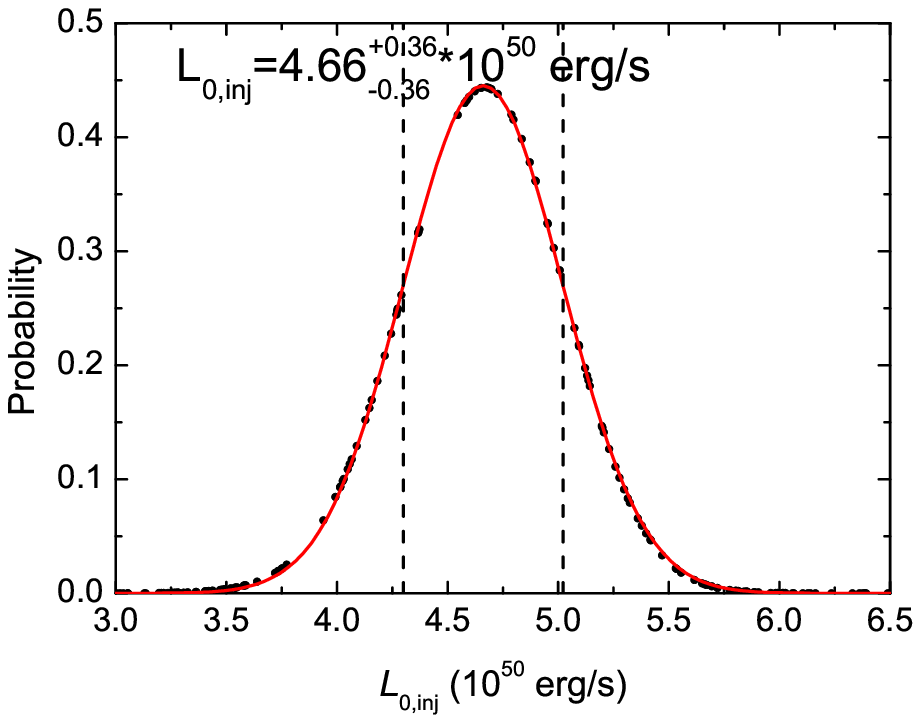}
\includegraphics[angle=0,width=0.3\textwidth]{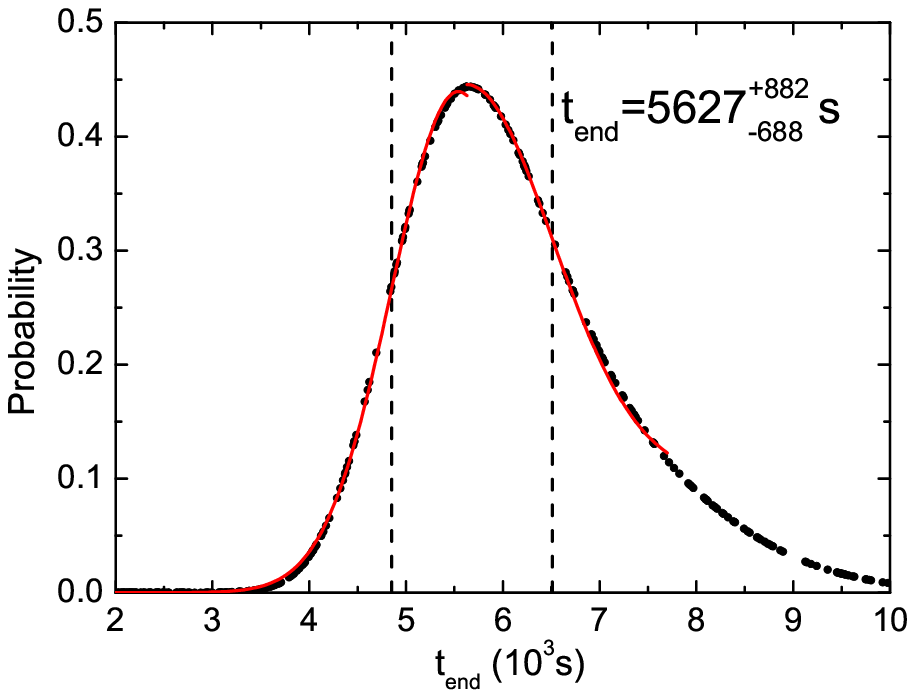}
\includegraphics[angle=0,width=0.3\textwidth]{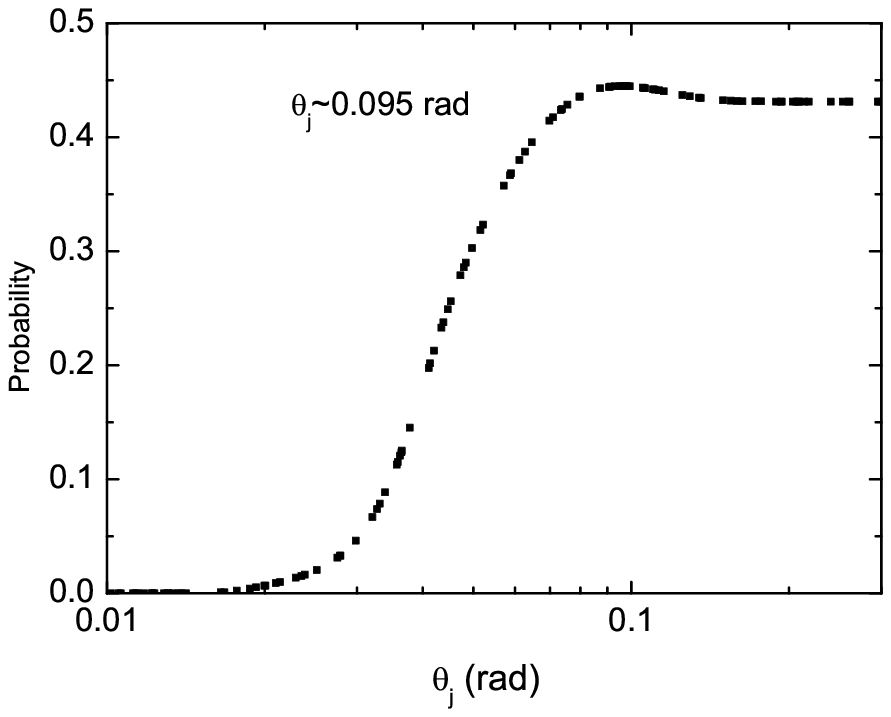}
\caption{Probability distributions of the afterglow model parameters along with our Gaussian fits ({\em solid} lines). The dashed vertical lines mark the $1\sigma$
standard deviations.\label{Model_parameter}}
\end{figure}%

 \begin{figure}[htbp]
 \centering
\includegraphics[angle=0,width=0.8\textwidth]{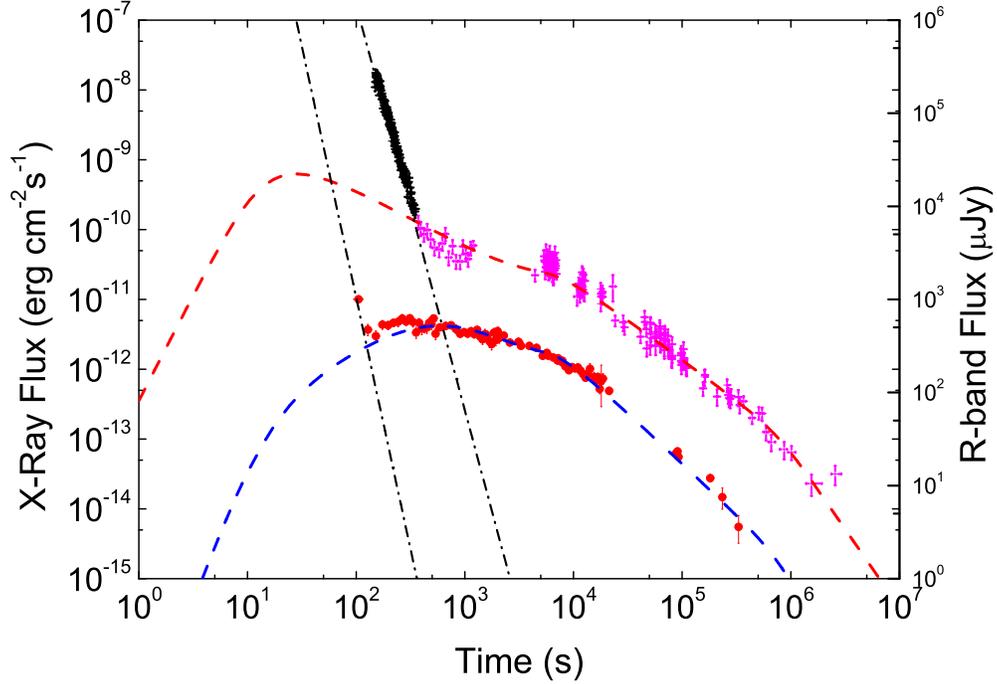}
\caption{Illustration of our best fit ({\em Dash} lines) to the afterglow data with the standard afterglow model incorporating with both the energy
injection and jet break effects via an approach of Monte Carlo simulations. The time intervals for our fits are $t>700$ seconds for the X-ray data and $t>140$ seconds for the optical data. Empirical fits with a single power-law ( {\em dash-dotted} lines) for data prior the time intervals are also shown. \label{Model_fitting}}
\end{figure}

\begin{figure}[htbp]
 \centering
\includegraphics[angle=0,width=0.45\textwidth]{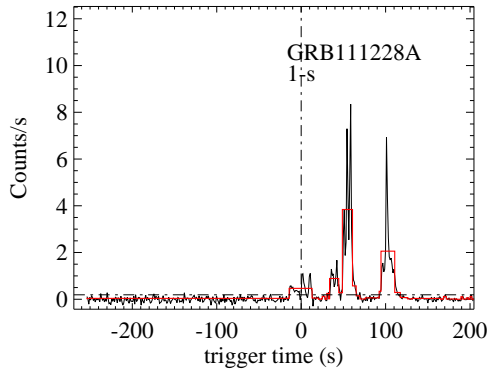}
\caption{BAT lightcurve of GRB 111228A along with the analysis results with the Beyasian block method (e.g., Hu et al. 2014).
\label{Prompt_LC}
}
\end{figure}

\begin{figure}[htbp]
 \centering
\includegraphics[angle=0,width=0.4\textwidth]{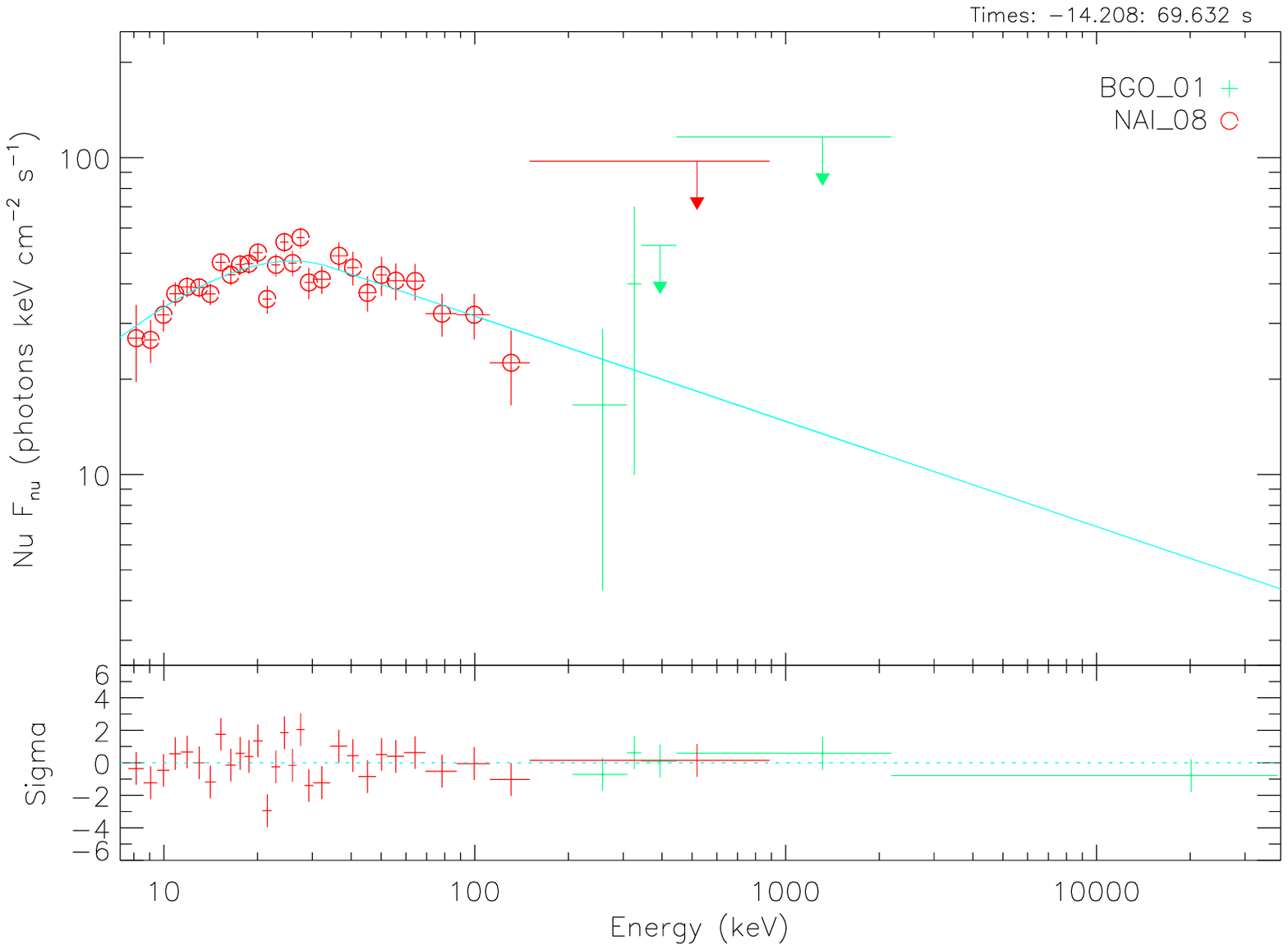}
\includegraphics[angle=0,width=0.4\textwidth]{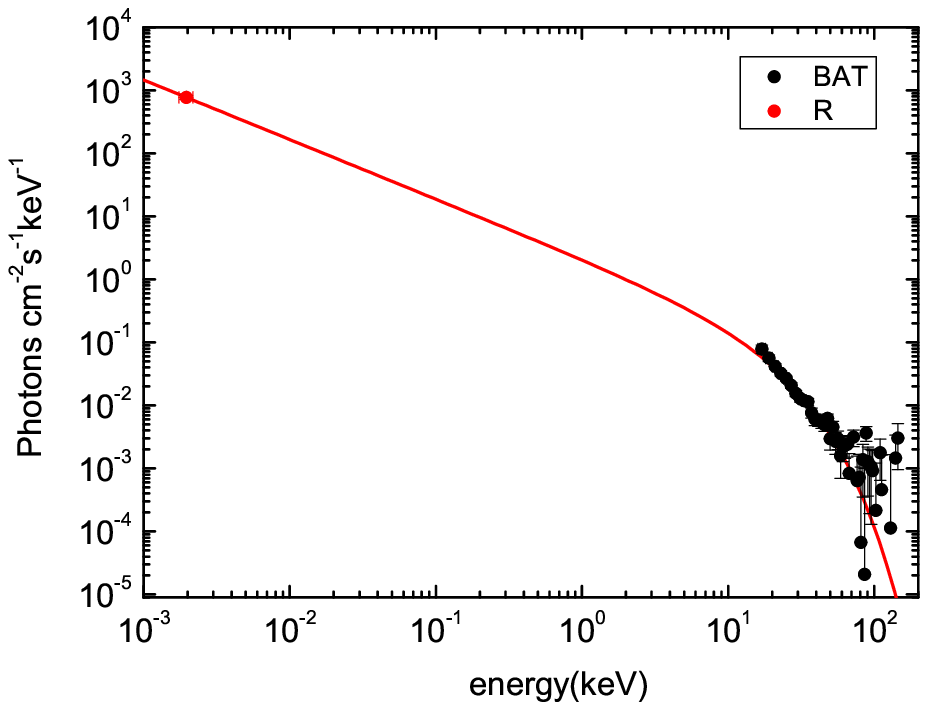}
\caption{{\em Left: }Time-integrated GBM spectrum of GRB 111228 along with our fit with the Band function. {\em Right: } Joint spectral fit to the spectrum energy distribution in the optical and BAT band for the time interval from 95 to 115 second post the BAT trigger (the of exposure time of our first optical data point and the last pulse of the prompt gamma-rays). The solid line is the extinction corrected spectrum derived from our fit with the Band function. Extinctions of our Galaxy and GRB host galaxy are considered. The extinction law of the host galaxy is adopted as the SMC and LMC laws. The host galaxy extinction is negligible in our fit.}
\label{Prompt_SED}
\end{figure}

\begin{figure}[htbp]
 \centering
\includegraphics[angle=0,width=0.8\textwidth]{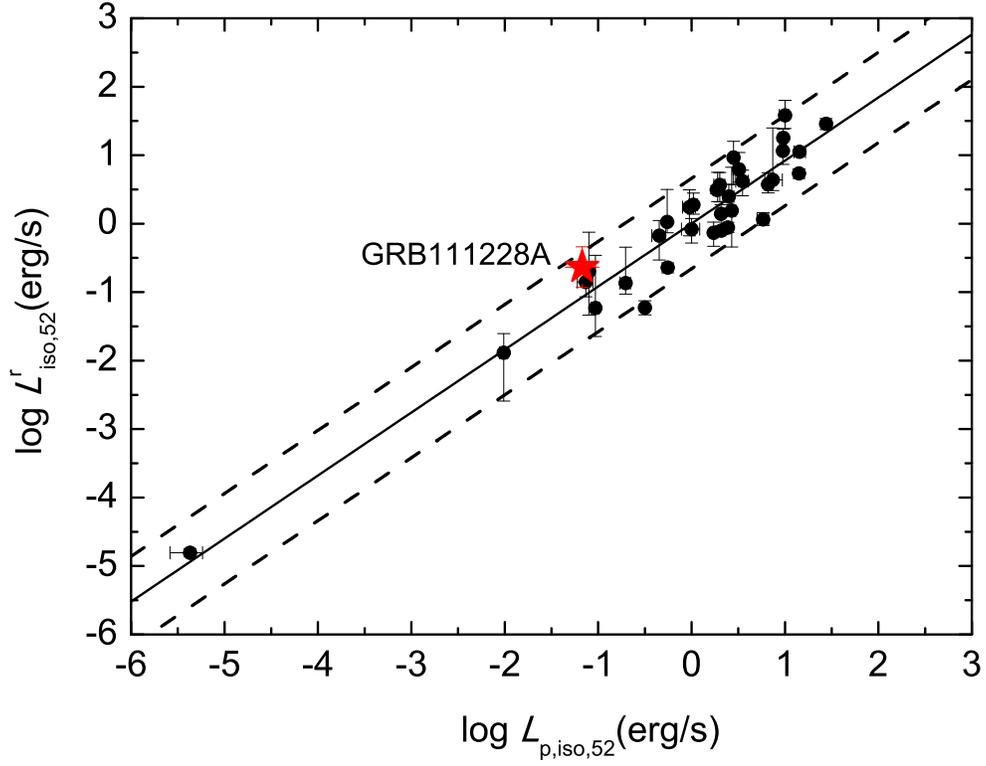}
\caption{Illustration of the satisfaction of GRB111228A ({\em the red star}) with the $L_{\rm iso}-E_{p,z}-\Gamma_0$ relation derived from other typical GRBs ({\em black dots}; taken from Liang et al. 2015). Lines are the best fit and the 3 $\sigma$ significance level.\label{Liso_Ep_gamma0}}
\end{figure}


 \begin{figure}[htbp]
 \centering
\includegraphics[angle=0,width=0.8\textwidth]{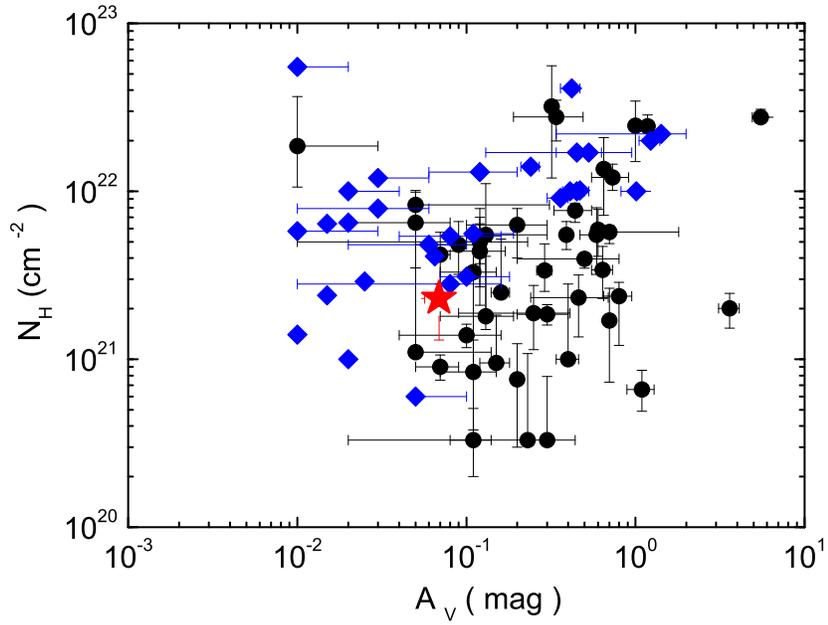}
\caption{Comparison of GRB 11228A ({\em the star} )with other typical long GRBs in the $N_{\rm H}-A_{\rm V}$ plane.
GRBs marked as {\em dots} are taken from Chen et al. (2014) and GRBs marked with {\em diamonds} are taken from Greiner et al. (2011).
\label{AVNH}}
\end{figure}%

 \begin{figure}[htbp]
 \centering
\includegraphics[angle=0,width=0.8\textwidth]{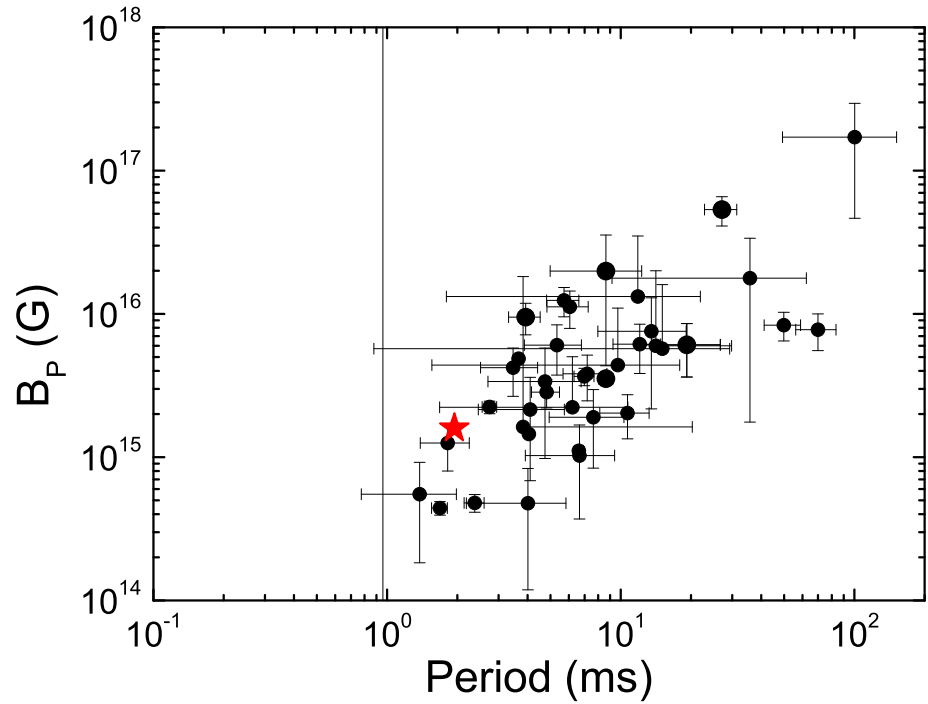}
\caption{Comparison of GRB 11228A ({\em the star} )with other typical magnetar candidates GRBs in the $B_p - P_0$ plane of beaming correction.
GRBs marked as {\em dots} are taken from L\"{u} \& Zhang (2014). The vertical
solid line is the breakup spin-period for a neutron star (Lattimer \& Prakash 2004).
\label{BpP0}}
\end{figure}

\end{document}